\documentstyle[twoside,fleqn,espcrc2,epsf]{article}

\def\3he{{$^3${\rm He}}}

\def\slD{\raise.15ex\hbox{$/$}\kern-.53em\hbox{$D$}}
\def\dsl{\raise.15ex\hbox{$/$}\kern-.57em\hbox{$\Delta$}}
\def\slp{{\raise.15ex\hbox{$/$}\kern-.57em\hbox{$\partial$}}}
\def\nsl{\raise.15ex\hbox{$/$}\kern-.57em\hbox{$\nabla$}}
\def\sla{\raise.15ex\hbox{$/$}\kern-.57em\hbox{$\rightarrow$}}
\def\slla{\raise.15ex\hbox{$/$}\kern-.57em\hbox{$\lambda$}}
\def\slb{\raise.15ex\hbox{$/$}\kern-.57em\hbox{$b$}}
\def\lnp{\raise.15ex\hbox{$/$}\kern-.57em\hbox{$p$}}
\def\lnk{\raise.15ex\hbox{$/$}\kern-.57em\hbox{$k$}}
\def\lnK{\raise.15ex\hbox{$/$}\kern-.57em\hbox{$K$}}
\def\lnq{\raise.15ex\hbox{$/$}\kern-.57em\hbox{$q$}}

\def\cM{{\cal M}}

\def\pmb#1{\setbox0=\hbox{$#1$}%
\kern-.025em\copy0\kern-\wd0
\kern.05em\copy0\kern-\wd0
\kern-.025em\raise.0433em\box0 }

\def\q2{{Q^2}}
\def\gtwid{\raise.3ex\hbox{$>$\kern-.75em\lower1ex\hbox{$\sim$}}}
\def\ltwid{\raise.3ex\hbox{$<$\kern-.75em\lower1ex\hbox{$\sim$}}}
\def\12{{1\over2}}
\def\part{\partial}

\def\low#1{\lower.5ex\hbox{${}_#1$}}

\def\psl{\raise.15ex\hbox{$/$}\kern-.57em\hbox{$\partial$}}
\def\partt{\raise.15ex\hbox{$\widetilde$}{\kern-.37em\hbox{$\partial$}}}

\def\topppageno1{\global\footline={\hfil}\global\headline
={\ifnum\pageno<\firstpageno{\hfil}\else{\hss\twelverm --\ \folio
\ --\hss}\fi}}
 
\def\toppageno2{\global\footline={\hfil}\global\headline
={\ifnum\pageno<\firstpageno{\hfil}\else{\rightline{\hfill\hfill
\twelverm \ \folio
\ \hss}}\fi}}

\def\prd#1{Phys.\ Rev.\ {\bf D#1}}
\def\plb#1{Phys.\ Lett.\ {\bf B#1}}

\def\et{{\it et al.}}

\def\nsection#1 #2{\leftline{\rlap{#1}\indent\relax #2}}

\def\prd#1{Phys.\ Rev.\ {\bf D#1}}
\def\plb#1{Phys.\ Lett.\ {\bf #1B}}

\def\dallas{Nucl.\ Phys.\ {\bf B} (Proc.\ Suppl.) {\bf 34} (1994)}

\def\melbourne{Nucl.\ Phys.\ {\bf B} (Proc.\ Suppl.) {\bf 47} (1996)}

\def\stlouis{Nucl.\ Phys.\ {\bf B} (Proc.\ Suppl.) {\bf 53} (1997)}

\newcommand{\AmS}{{\protect\the\textfont2
  A\kern-.1667em\lower.5ex\hbox{M}\kern-.125emS}}
\newcommand{\fB}{$f_B$}
\newcommand{\fBs}{$f_{B_s}$}
\newcommand{\fD}{$f_D$}
\newcommand{\fDs}{$f_{D_s}$}

\title{Heavy-light decay constants 
from Wilson and static quarks}

\author{ C.~Bernard,\hskip-0.03in
\address{{\vskip-0.10in{\hskip 0.07in Department of Physics, Washington
University, St.~Louis, MO 63130, USA}}} 
\thanks{presented by C.\ Bernard at {\it Lattice 97}, Edinburgh, July 22--26, 1997}
T.~DeGrand,\hskip-0.03in
\address{Physics Department, University of Colorado, Boulder, CO 80309, USA} %
C.~DeTar,\hskip-0.03in
\address{Physics Department, University of Utah, Salt Lake City, UT 84112, USA}
Steven~Gottlieb,\hskip-0.03in
\address{Department of Physics, Indiana University, Bloomington, IN 47405, USA}
U.~M.~Heller,\hskip-0.03in
\address{SCRI, Florida State University, Tallahassee, FL 32306-4130, USA} 
J.~Hetrick,\hskip-0.03in$\,\null^{\rm a}$\thanks{present address: Department of
Physics, University of the Pacific, Stockton, CA 95211, USA} 
C.~McNeile,\hskip-0.03in$\,\null^{\rm c}$
K.~Rummukainen,\hskip-0.03in
\address{Universit\"at Bielefeld, Fakult\"at f\"ur Physik, Postfach 100131, D-33501 Bielefeld, Germany} 
R.~Sugar,\hskip-0.03in
\address{Department of Physics, University of California, Santa Barbara, CA
93106, USA} 
D.~Toussaint,\hskip-0.03in
\address{Department of Physics, University of Arizona, Tucson, AZ 85721, USA} 
and M.~Wingate$\,\null^{\rm b}$
} 

\begin{document}

\begin{abstract}
MILC collaboration results for \fB, \fBs, \fD, \fDs and their ratios
are presented.  These results are still preliminary, but the analysis
is close to being completed.  Sources of systematic error,
both within the quenched approximation and from quenching itself,
are estimated.
We find, for example,
$f_B=153\ \pm 10\ {}^{+36}_{-13} \ {}^{+13}_{ -0}\ {\rm MeV}$,
and $f_{B_s}/f_B  = 1.10\ \pm 0.02\ {}^{+0.05}_{ -0.03} \ {}^{+0.03}_{ -0.02}$,
where the errors are statistical, systematic (within the quenched
approximation), and systematic (of quenching), respectively.
The extrapolation to the continuum and the
chiral extrapolation are the largest sources of error.  Present
central values are based on linear chiral extrapolations;
a shift to quadratic extrapolations would 
raise $f_B$ by $\approx\!20$ MeV and make the error within the
quenched approximation more symmetric.
\vspace{-10pt}

\end{abstract}

\maketitle

The MILC collaboration is continuing its 
pro\-gram~\cite{milc-older,milc-lat96,milc-tsukuba,milc-b20}
 of calculating the
decay constants of heavy-light pseudoscalar mesons.
The computations use Wilson light quarks and Wilson and static
heavy quarks.  We work on both quenched lattices, with a wide
range of lattice spacings, and $N_F\!=\!2$ dynamical staggered lattices.
Table~\ref{tab:lattices} gives the lattice parameters.

\begin{table}
\caption{Lattice parameters.  Lattices F, G, and L--P use
variable-mass Wilson valence quarks and
two flavors of fixed-mass staggered
dynamical fermions;
all other runs use quenched
Wilson quarks. 
Lattice G was generated by HEMCGC; lattice F, by the Columbia group. }
\label{tab:lattices}
\centering
\begin{tabular}{cccc} \hline
name& $\beta\  (am_q)$ &size &\# configs. \\
\hline
\vrule height 10pt width 0pt A  & 5.7 & $8^3\times 48$ & 200 \\
\hline
\vrule height 10pt width 0pt B  & 5.7 & $16^3\times 48$ & 100 \\
\hline
\vrule height 10pt width 0pt E& $ 5.85$&  $12^3 \times 48$&  100 \cr
\hline
\vrule height 10pt width 0pt Q  & 6.0 & $12^3\times 48$ & 235\\
\hline
\vrule height 10pt width 0pt C  & 6.0 & $16^3\times 48$ & 100\\
\hline
\vrule height 10pt width 0pt D  & 6.3 & $24^3\times 80$ & 100\\
\hline
\vrule height 10pt width 0pt H& $ 6.52$&  $ 32^3 \times 100$& 60 \cr
\hline
\vrule height 10pt width 0pt F& $ 5.7\ (0.01)$&   $16^3 \times 32$&    49 \cr
\hline
\vrule height 10pt width 0pt G& $ 5.6\ (0.01)$&   $16^3 \times 32$&    200 \cr
\hline
\vrule height 10pt width 0pt L& $ 5.445\ (0.025)$&  $ 16^3 \times 48$&    100
\cr
\hline
\vrule height 10pt width 0pt N& $ 5.5\ (0.1)$&  $ 24^3 \times 64$&    100 \cr
\hline
\vrule height 10pt width 0pt O& $ 5.5\ (0.05)$&  $ 24^3 \times 64$&    100 \cr
\hline
\vrule height 10pt width 0pt M& $ 5.5\ (0.025)$&  $ 20^3 \times 64$&    100(200) \cr
\hline
\vrule height 10pt width 0pt P& $ 5.5\ (0.0125)$&  $ 20^3 \times 64$&    199
\cr
\hline

\hline
\vspace{-30pt}
\end{tabular}
\end{table}

A major improvement in the past year has been the completion
of dedicated runs to determine the static-light decay
constants on lattices A,B,Q,E,G,L,N,O,M,P.
These runs use a multi-source technique, with relative wavefunctions  
taken from \cite{wavefunctions}.
On lattices  C,D,H,F,G --- generally the smaller physical
volumes --- 
we get acceptable static-light data as a simple
by-product of the Henty-Kenway hopping parameter expansion \cite{henty}
used for the heavy Wilson quarks. On lattice G, where both
methods are available, the results are consistent. 

A second improvement has been the calculation,
following
the approach of \cite{lepmac}, of the 
scale $q^*$ of the coupling in the 
perturbative renormalization
constant $Z_A$ of
the axial current. 
For propagating Wilson
quarks, the result is, after
tadpole improvement, $q^*=2.32/a$ \cite {bgm}.
Mass dependent effects are not included at this point.
We estimate the systematic error of the renormalization
by changing $q^*$ by a factor of 2 and reanalyzing.
The error is rather small ($\ltwid 3\%$). 

Currently, we find (in MeV):
\vspace{-4pt}
\begin{eqnarray*}
f_B \!=\! 153 (10) ({}^{+36}_{-13})(  {}^{+13}_{ -0});\ 
f_{B_s}\!\!\!\! \!&=&\!\!\!\!\! 164 (9) ({}^{+47}_{ -13})(  {}^{+16}_{ -0})\\
f_D \!=\! 186(10) ({}^{+27}_{ -18} )({}^{+9}_{ -0});\ \ 
f_{D_s} \!\!\!\!\!&=&\!\!\!\!\! 199 (8) ({}^{+40}_{ -11} )({}^{+10}_{ -0}).
\vspace{-4pt}
\end{eqnarray*}
The errors are statistical (plus ``fitting''),
systematic (within the quenched
approximation), and systematic (of quenching), respectively.
For the ratios, we get:
\vspace{-1pt}
\begin{eqnarray*}
f_{B_s}/f_B \!\!&=&\!\! 1.10 (2) ({}^{+5}_{ -3})  ({}^{+3}_{ -2})\quad
\phantom{f_{B_s} = 164 (9) }\\
f_{D_s}/f_D\!\! &=&\!\! 1.09 (2) ({}^{+5}_{ -1})  ({}^{+2}_{ -0})\quad
\phantom{f_{B_s} = 164 (9) }\\
f_{B}/f_{D_s}\!\! &=&\!\! 0.76 (3)({}^{+7}_{ -4})({}^{+2}_{ -0}).\quad
\phantom{f_{B_s} = 164 (9) }
\vspace{-8pt}
\end{eqnarray*}
A discussion of the most important 
sources of systematic errors follows.

$\bullet$ Chiral Extrapolation.  The proper functional form to use
in extrapolating physical quantities to the chiral limit is not clear.
For example, although lowest order chiral perturbation theory
predicts that $m^2_\pi$ is a linear function of quark mass,
we observe \cite{milc-lat96,milc-tsukuba,milc-b20}
small but
significant deviations from linearity. 

These deviations 
could be due to unphysical effects such as
the finite lattice spacing or volume.
In addition,
the curvature can be changed significantly, and sometimes
made negligible, 
by shifting the fit range (in t) on the
individual propagators.  
Even the more
``physical'' cause (chiral logs or higher order 
analytic terms) are a source of possible
spurious effects
because quenched chiral logs are in general different from those in the
full theory
\cite{qchpt}.  

For these reasons, we presently
fit quantities like $m^2_\pi$ to their lowest order chiral
form, despite the poor confidence levels.  The systematic error
is estimated by repeating the analysis with quadratic
(constrained) fits.
This error is $\le\!10\%$ for decay constants
on all quenched data sets
used to extrapolate to the continuum; usually it is $\ltwid 5\%$.  (After
extrapolation to the continuum, the error is larger: $7\%$ to $15\%$.)

Our reasons for choosing linear chiral fits for the
central values are somewhat subjective, and it is possible that we
will switch to quadratic fits in the final version of this work.  
To help us make the choice, we are studying a large sample of 
quenched lattices
at $\beta\!=\!5.7$, with volumes up to $24^3$ \cite{spectrum}.  
On this sample we
have six light quark masses (as opposed to three for each of the lattices
used for the heavy-light computation) and have light-light
mesons with nondegenerate  as
well as degenerate quarks.

If a switch to quadratic fits were to be made now, it
would raise the central values of \fB, \fBs, \fD\ and \fDs\ by
23, 19, 13, and 14 MeV, respectively.  The systematic
error within the quenched approximation would then become much more symmetric,
with the continuum extrapolation the dominant positive error and the
chiral extrapolation the dominant negative one. 

$\bullet$ Heavy-Quark Interpolation.
Having static-light results on all lattices allows us to find decay
constants for physical $B$ mesons by interpolation between heavy-light
and static-light data,
rather than extrapolation from the former.
The interpolating fits  have good 
confidence levels on all our data sets, reassuring us that the
procedure \cite{ekm} we use for the heavy-light data is reasonable.  

One estimate of the systematic error of this approach 
is obtained by comparing decay constants computed with two different 
mass ranges of propagating quarks: ``lighter heavies,'' 
(mesons 1.25 to 2 GeV) and ``heavier heavies,''
(mesons 2 to 4 GeV).  
The difference is less than 1\% at the
three weakest quenched couplings ($\beta \!=\!6.0, 6.3, 6.52$),
less than 5\%
over all lattices, and less than 4\% after linear extrapolation of all quenched 
lattices to $a\!=\!0$.

For quenched lattices A,B,E, the new static-light results
produce only small changes from that reported previously \cite{milc-lat96}.
However, on the large $N_F\!=\!2$ lattices, including the static point raises $f_B$ by
about one old (statistical) standard deviation, and reduces the statistical
error (and the difference between using heavier-heavies and lighter-heavies)
by about 50\%.  

$\bullet$ Extrapolation to the Continuum.
For any physical quantity Q computed here, we expect 
$ Q(a) = Q_{a=0}(1 + a\cM_1 + \cdots )$.
In practice, we find the slope to be large
for the decay constants ($\cM_1 \sim 300$--$650$ MeV), with \fBs\ the
worst offender.  This leads to large extrapolation errors
($\sim\!12$--$27\%$).  The ratios of decay constant
are  much better behaved, with
$\cM_1\sim 100$ MeV and an error of $\sim\!4$--$5\%$.

\begin{figure}[htb] 
\vspace{-44pt}
\epsfxsize=1.0 \hsize
\epsffile{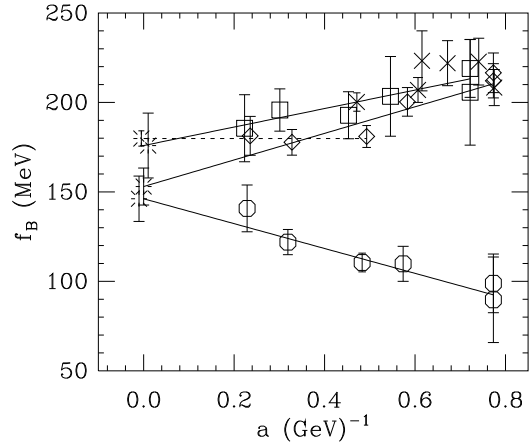}
\vspace{-28pt}
\caption{ Results for \fB\ as a function of lattice spacing.
The diamonds are quenched approximation points.
Octagons and squares 
are used for estimating systematic errors --- see text.
Crosses are from $N_F\!=\!2$ lattices but are otherwise computed
like the diamonds.
}
\vspace{-19pt}
\label{fb}
\end{figure}

Figure~\ref{fb} shows several fits of \fB\ {\it vs.}\ $a$ used
to estimate the two largest sources
of systematic error.  The central value is obtained from
a linear fit to all the diamonds, which in turn use linear chiral
fits, a lattice scale set by $f_\pi$, and the ``EKM'' corrections \cite{ekm}.
The error of the continuum extrapolation is
estimated by comparing the central value with the results of 
a constant fit to the three diamonds with smallest values of $a$,
a linear fit to the octagons (which use ``lighter heavies'' and
no EKM corrections), 
and two other similar types of fits (not shown).  The 
continuum extrapolation
error is defined as the largest of these four differences and is
in practice almost always given by the first difference.
The difference of the extrapolation
of the squares (which have a quadratic chiral extrapolation) and the
central value determines the chiral extrapolation error.

$\bullet$ Quenching Effects.  We have repeated our com\-pu\-ta\-tions 
on several $N_F$=$2$ dy\-nam\-i\-cal
fer\-mion lattices (crosses in Fig.~\ref{fb}).
Such computations
are not yet ``full QCD''  because (1) the virtual quark mass
is fixed and not extrapolated to the chiral limit,
(2) the $N_F\!=\!2$
data is not yet
good enough to extrapolate to $a\!=\!0$, and (3) we have two light flavors,
not three.  Thus the $N_F\!=\!2$ simulations are used at this point only
for systematic error estimation.  See \cite{milc-b20}  for details.

We thank CCS (ORNL), 
Indiana University, 
SDSC,
PSC, CTC, MHPCC,  CHPC (Utah), and Sandia Natl.\ Lab.\
for computing resources.
This work was supported in part by the DOE and NSF.

\end{document}